\documentclass[reprint,superscriptaddress,twocolumn,aps,prb,showpacs]{revtex4-1}
\usepackage{graphicx}
\usepackage{amsbsy,amssymb,amsmath,bm,ulem}

\usepackage{color}

\graphicspath{{../pic/}}

\begin{document}
\title{Mechanism for flux guidance by micrometric antidot arrays in superconducting  films}
\author{J. I. Vestg{\aa}rden}
\affiliation{Department of Physics, University of Oslo, P. O. Box
1048 Blindern, 0316 Oslo, Norway}
\author{V. V. Yurchenko}
\affiliation{Department of Physics, University of Oslo, P. O. Box
1048 Blindern, 0316 Oslo, Norway}
\author{R. W{\" o}rdenweber}
\affiliation{Peter Gr{\"u}nberg Institute and JARA-Fundamentals of Future Information Technology, Forschungszentrum J{\"u}lich, 52425 J{\"u}lich, Germany}
\author{T. H. Johansen}
\affiliation{Department of Physics, University of Oslo, P. O. Box
1048 Blindern, 0316 Oslo, Norway}
\affiliation{Institute for Superconducting and Electronic Materials,
University of Wollongong, Northfields Avenue, Wollongong,
NSW 2522, Australia}
\affiliation{Centre for Advanced Study at The Norwegian Academy
of Science and Letters, Drammensveien 78, 0271 Oslo, Norway}

\begin{abstract}
A study of magnetic flux penetration in a superconducting film patterned with arrays 
of micron sized antidots (microholes) is reported.
Magneto-optical imaging (MOI) of a YBa$_2$Cu$_3$O$_x$ film shaped as a long strip
with perpendicular antidot arrays revealed both strong guidance of flux, and
at the same time large perturbations of the overall flux penetration and 
flow of current. These results are compared with a numerical flux creep simulation 
of a thin superconductor with the same antidot pattern.
To perform calculations on such a complex geometry, an efficient numerical scheme
for handling the boundary conditions of the antidots and the nonlocal 
electrodynamics was developed. The simulations reproduce essentially all features 
of the MOI results. In addition, the numerical results give insight into all other 
key quantities, e.g., the electrical field, which becomes extremely large in the narrow 
channels connecting the antidots.
\end{abstract}


\maketitle 

\section{Introduction}

The  motion of magnetic flux in type-II superconducting films can to a
large extent be controlled by introduction of 
artificial micro- and nano-structures, such as antidots (holes),\cite{baert95,eisenmenger01,
pannetier03} magnetic dots,\cite{martin97,gheorghe08}
thickness modulations,\cite{ivanchenko83-2,he09}
grain boundaries,\cite{polyanskii96,bartolome07} slits,\cite{baziljevich96-2} or
magnetic domain walls in superconductor/ferromagnet
hybrids.\cite{goa03,vlasko-vlasov08-2} Such structures are key building blocks for 
successful realization of fluxonics devices like vortex ratchets, pumps and 
lenses etc.\cite{villegas03, desilva06-prb,  wambaugh99}
It is known that when antidots are sufficiently small they 
can become pinning sites for the vortices.\cite{mkrtchyan72}  
This type of pinning is usually noticable only close to the critical
temperature $T_c$, and can be observed, e.g.,  as pronounced matching between 
the vortex density and the underlying  lattice of antidots.\cite{moshchalkov98, silhanek03}
It was also demonstrated that certain patterns can be used to 
reduce noise due to vortex motion in SQUIDs,\cite{wordenweber02} and 
quite recently in superconducting microwave resonators.\cite{bothner11}

Of technological as well as fundamental interest is also realizations of flux guidance, i.e., 
how to achieve directed and controlled motion of the magnetic flux.  
When the pinning by the antidots 
is strong compared to the intrinsic  pinning of the material,  guidance is 
effectuated by the more mobile interstitial vortices.\cite{reichhardt09}
Conversely, when the intrinsic pinning is strong, flux moves most easily inside
the holes, and arrays of antidots should enhance the flux penetration.\cite{vestgarden08}
It has been shown that by introducing arrays 
of dots\cite{villegas03-prb} or antidots\cite{silhanek03,wordenweber04} 
the vortices can be guided away from the direction given by the Lorentz force of
an applied current.  Similarly, a periodic arrangement of antidots
can cause effective flux drainage of a sample in the descending
branch of a magnetic field ramp.\cite{crisan05} 
It was also found that when the antidot lattice breaks the symmetry of 
the  overall sample shape, or when the antidots have nontrivial shapes,
the critical current density can become anisotropic\cite{pannetier03, gheorghe06}   
and the flux motion can be enhanced in unexpected directions.\cite{tamegai10}

Whereas superconducting films containing complex arrangements of
antidots can today be readily produced using, e.g., optical
lithography, the theoretical modelling of the local and global
features of the flux dynamics in such systems is challenging.
In essence, this is due to the nonlocal nature of the electrodynamics
in two-dimensional samples subjected to a perpendicular 
magnetic field. In this case the critical-state is strongly modified since 
at any applied field there will be currents flowing over the
whole sample area, including 
regions in the flux-free Meissner-state.\cite{norris69, brandt93, zeldov94}
Also numerical simulations of the flux dynamics 
are a lot more computationally demanding 
in films compared to bulk.\cite{brandt95,loerincz04,vestgarden11}
Additional complications appear
in a film with antidots, as the shielding currents then meet
constrictions, and the flow must adapt to the available and often 
narrow bridges of superconducting material between the antidots.  
Hence, the current density quickly rises to the critical value, causing  major
rearrangements of both the current flow and the flux distribution.\cite{vestgarden08} 

Magneto-optical imaging (MOI) is a unique experimental tool for
observing the nontrivial redistributions created by patterning of
superconducting films. The technique allows direct observation of the
flux density over length scales ranging from the entire
sample size and down to the size of individual antidots. 
In spite of large previous efforts, surprisingly few investigations were
carried out on films with simple antidot patterns. 
In this work we present results from MOI experiments together with theoretical
modelling of the flux and current behavior in a thin superconducting strip
containing only a  few linear arrays of antidots. The experiments were performed 
using a film of YBa$_2$Cu$_3$O$_x$ (YBCO) cooled 
far below $T_c$ to be in the regime where flux pinning by the antidots is negligible.


The paper is organized as follows.
In Sec.~II we outline the current flow modification caused by an antidot array 
as expected from the critical state model.  
Section~III presents results from our MOI experiments. 
Section~IV describes our method for numerical simulations of the electrodynamics, and  
the results  are presented and discussed in Sec.~V.  Section~VI gives a summary.

\section{Critical current flow lines}

\begin{figure}
  \centering
  \includegraphics[width=8cm]{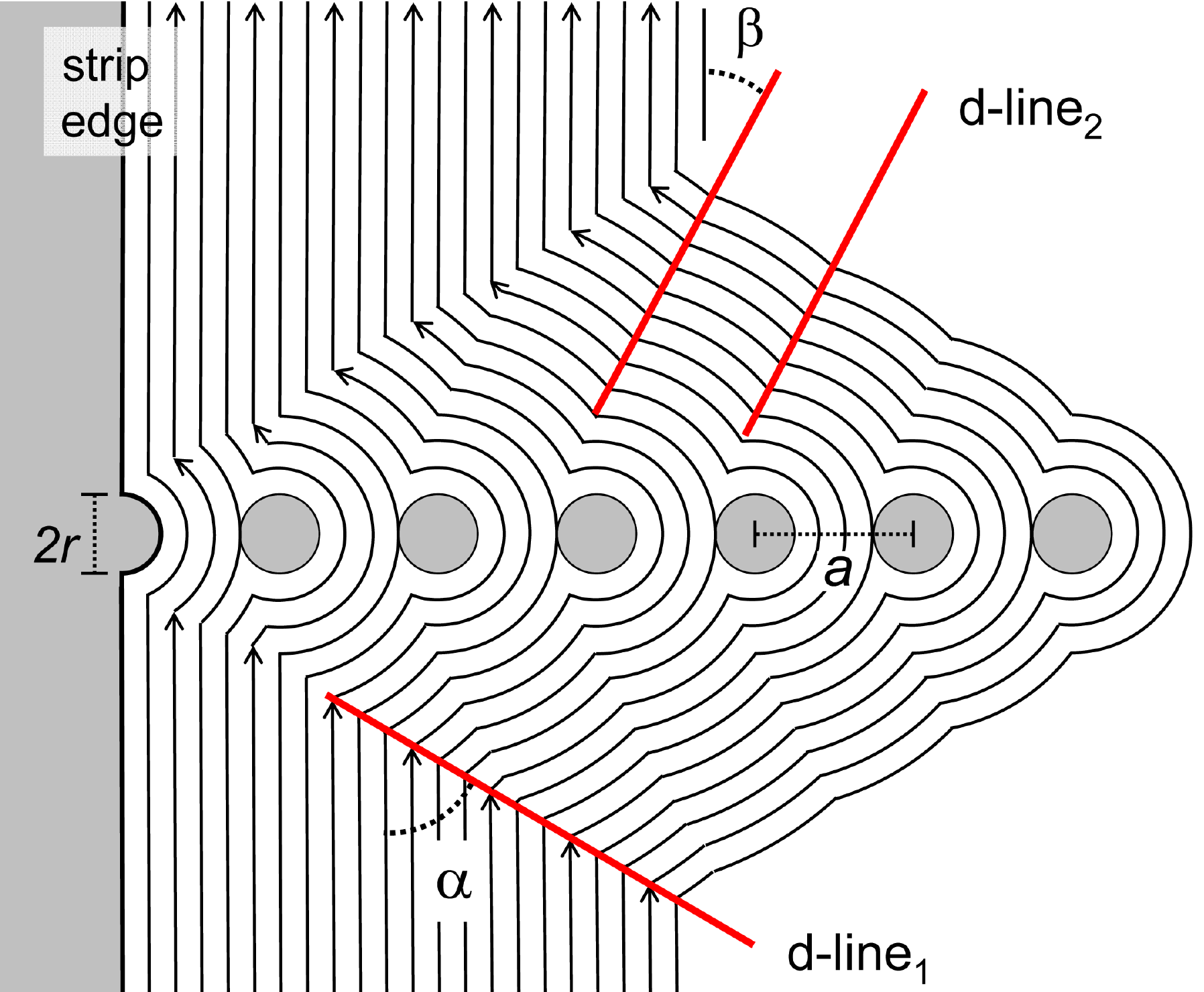} 
  \caption{
    \label{fig:bean}
    (Color online)
    Construction of the critical current streamlines near a linear array 
    of equidistant circular antidots. The pattern follows from the Bean critical-state model 
    for a sample with a straight edge. Note that streamlines of the sub-critical
 currents flowing in the Meissner-state part of the strip are not included.
  }
\end{figure}

Before presenting the experimental results 
we consider the predictions of the Bean critical-state model for
a type-II superconducting strip with a linear array of equally spaced
antidots crossing from one side to the other, see Fig.~\ref{fig:bean}.
The antidots have radius $r$ and center-to-center distance $a$. An
external magnetic field, $H_a$, is applied perpendicular to the strip
plane, and we assume the superconductor initially contains no flux and
current.

As the field is gradually  increased magnetic flux will penetrate from the edge
and shielding currents will begin to flow.  In areas
containing flux, the critical state is formed and the current density
has the critical magnitude $j_c$,
while in the unpenetrated Meissner-state region the
currents have a subcritical density.  
To draw the current stream lines in the critical-state region we use the
simple rule following from the Bean critical-state model, namely that the
magnitude of the current density is constant, and thus represented by
a set of equidistant stream lines with spacing inversely proportional 
to $j_c$.  The construction starts by drawing
continuous lines that follow the external perimeter of the sample, and
when reaching a hole the current must flow around it. This scheme results
in the flow pattern shown in the Fig. \ref{fig:bean}, corresponding to a partly
penetrated state. The closure of each streamline takes place in parts of
the sample not included in the figure.

The figure shows that the presence of the antidot array deforms the
critical current in a large region with a near rhombic shape.  A line
denoted d-line$_1$ marks where the direction of the current flow
suddenly begins to deviate from being parallel to the edge.  Using the
principle of current flow continuity one finds that the line makes an
angle $\alpha$ with the edge, given by
\begin{equation}
  \cos \alpha = \sqrt{r/a} \  .
  \label{alpha}
\end{equation}
Evidently, such a line exists on both sides of the array, 
and the pair defines one half of the rhombic area. 

In addition to the large perturbation of the current flow,
the construction also results in a fine structure in the flow pattern.
Due to the shape of the antidots the stream lines inside the rhombic 
area consist of a sequence of circular arcs. As seen from the figure, 
each antidot creates its own pair of lines along a direction defined 
by the cusped joints of two
arcs.  These lines, denoted d-line$_2$,  become straight a short distance 
from the array, and one finds that the angle, $\beta$, they make
 with the strip edge is the same for all of them, and given by
\begin{equation}
  \cot \beta = 2\frac{\sqrt{r(a-r)}}{a-2r} \  .
  \label{beta}
\end{equation}

Related to this construction is a previous analysis of the current flow
in the case of having a weak link across the superconducting 
strip.\cite{polyanskii96} It was assumed that inside the weak link the 
critical current density 
is uniformly reduced by a factor $j_b/j_c$.
Also that construction resulted in a rhombic area where the current 
has changed direction, and the angle $\alpha$ was found to be
given by $\cos 2\alpha = -(j_b/j_c)$. Indeed, this is equivalent to
Eq.~\eqref{alpha} with the transparency of the antidot array, $(a-2r)/a$, 
replaced by the current ratio in the weak link case.

\begin{figure}[bb]
  \centering 
  \includegraphics[width=\columnwidth]{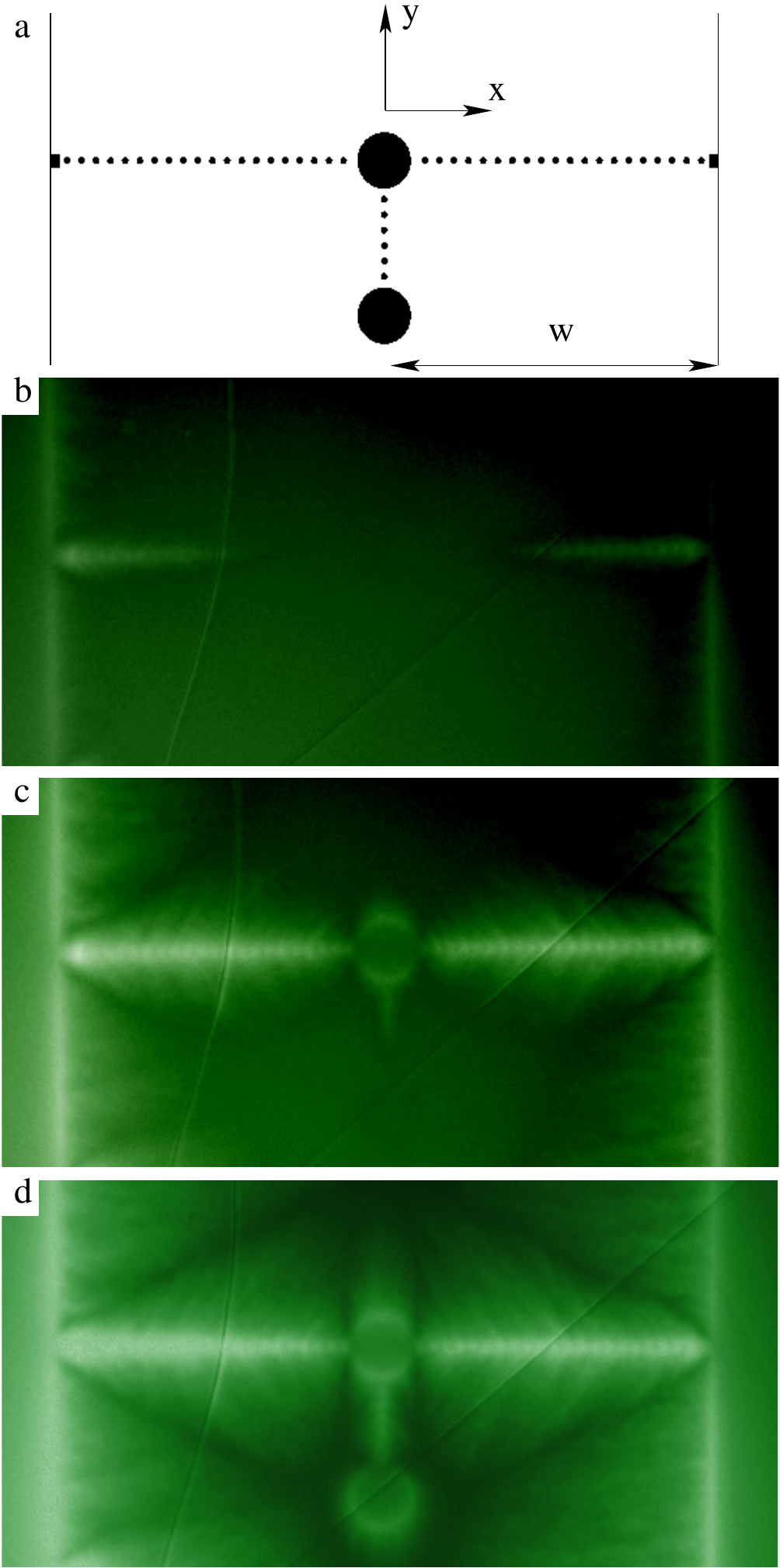} 
  \caption{
    \label{fig:mo}
    \label{fig:sample}
    (Color online)
    Sketch of the antidot arrangement in the superconducting 
    sample, which has an overall shape as a long thin strip in the $xy$-plane (a).
    The flux distribution, $B_z$, in the YBCO film 
seen by MOI and recorded at applied fields of 2.4 mT, 5.7 mT and 10 mT, respectively (b-d).
  }
\end{figure}

\section{Experiment}
A 150~nm thick YBCO film was produced by magnetron sputtering on a
r-cut sapphire substrate. The sample was shaped using optical
lithography and ion beam etching into a long strip 0.5~mm in width,
and with an arrangement of antidots as shown in Fig.~\ref{fig:sample}a. 
The antidots having radius $r= 2~\mu$m form a linear array  with period
$a=10~\mu$m. Along the center line there are two larger antidots of radius $20~\mu$m,
connected with an array of 6 antidots, also of radius $r= 2~\mu$m.
The arrangement was motivated by a possibility to study flux guidance along linear antidot 
arrays both perpendicular and  parallel to the strip edges.
The large holes are included to serve as reservoirs for the incoming flux.

Magneto-optical imaging  of the sample was performed using a 
bismuth substituted ferrite garnet film
with in-plane magnetization as Faraday rotating sensor.\cite{helseth01}
Images of the flux distribution were recorded through a polarized light microscope
using crossed polarizers. In this way the image brightness represents 
the magnitude of the flux density.
For details of the setup, see Ref.~\onlinecite{goa03-2}.

\begin{figure}[t]
  \centering
  \includegraphics[width=75mm]{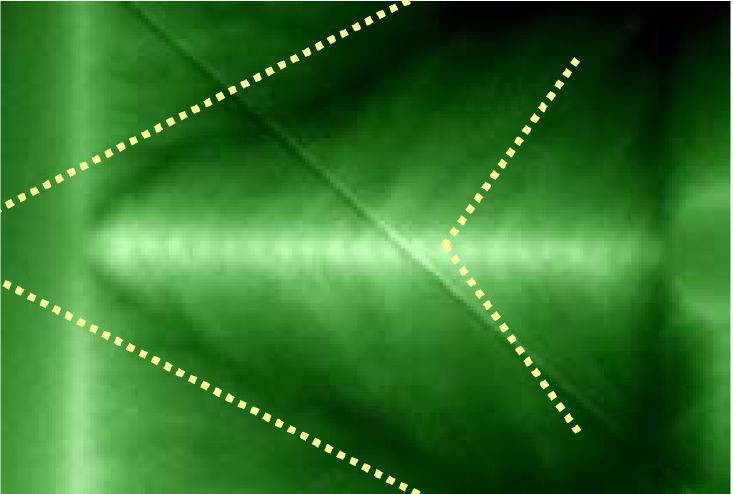} 
  \caption{
    \label{fig:mo-detail}
    (Color online)
    Details of Fig.~\ref{fig:mo} showing the d-line pattern from a 
    row of antidots. The straight d-line$_1$ and d-line$_2$ are drawn
    as a guide to the eye with angles corresponding to Eqs.~\eqref{alpha} and \eqref{beta}
calculated for the geometrical characteristics of the antidot array in the YBCO film.
  } 
\end{figure}

Shown in Fig.~\ref{fig:mo}b-d  is the flux distribution, $B_z$, in the
patterned part of the sample as the applied field is slowly ramped up
 to $\mu_0 H_a = 10$~mT
after an initial zero-field-cooling to 50~K.
As typical for thin films, the rim of the strip appears bright, thus
showing piling up of the field that is expelled by the superconductor.
Already at a field of 2.4~mT (panel b) the linear arrays perpendicular to the
edges are clearly visible, giving direct evidence that under these conditions the antidots are guiding the flux rather than pinning the vortices. 
When the field becomes 5.7~mT (panel c) guided flux reaches the large hole
in the center of the strip, and interestingly, the flux continues its motion 
out of the hole being guided by the antidot array oriented parallel to the edges.
In panel d, one sees that at 10~mT the parallel guidance  has filled
also the second large hole with flux well before the strip is fully penetrated.

Comparing the MOI results with the current flow pattern outlined
 in Fig.~\ref{fig:bean}, 
note in the image of Fig.~\ref{fig:mo}d  that
 four straight dark lines form a diamond shape 
where the two transverse antidot arrays make up the long diagonal. 
These dark lines 
correspond to discontinuity lines of the type marked as  
 d-line$_1$ in the critical-state construction.
 From Eq.~\eqref{alpha} and
the dimensional characteristics of the present antidot array 
 the angle $\alpha$ should  be given by $\cos \alpha =
1/\sqrt{5}$, or $\alpha = 63^{\circ}$.  Shown in
Fig.~\ref{fig:mo-detail} is a close-up view of $B_z$ near the antidot
array superimposed with a pair of dotted lines tilted by $\alpha =
63^{\circ}$, demonstrating an excellent quantitative agreement. Also
the fact that the diamond-feature shows up as dark, i.e., has a very low
flux density, follows naturally from the construction in
Fig.~\ref{fig:bean} since the sharp clock-wise turning of the current
near the d-line$_1$ provides an additional local shielding.

The fine structure of $B_z$ in the region around the antidot arrays is
only faintly visible. This is not surprising considering the d-line$_2$ in Fig.~\ref{fig:bean},  
where the turning of the currents at the cusps is gradually reduced away from the antidots.  
Nevertheless, one can clearly see in Fig.~\ref{fig:mo-detail} traces of a
fishbone-pattern. The angle
$\beta$ should according to Eq.~\eqref{beta} be $\beta =
37^\circ$, and Fig.~\ref{fig:mo-detail} includes as a guide to the eye a
pair of dotted lines having this angle. One finds a quite nice
agreement with the streaky features visible in the magneto-optical
image on both sides of the antidot array.

Note that also the observed high flux density along the array of
antidots can be readily understood from the current flow construction
in Fig.~\ref{fig:bean}. As the current flows past the constricted
region the stream lines make a very sharp turn in the
counter-clockwise direction. This curvature enhances flux density of
the same polarity as the one entering from the edge, thus creating an
effective flux guidance well beyond the overall flux penetration front
in the strip.

Although many features of the MOI results can be understood from the
critical-state considerations above, this picture is far from
complete.  In particular, the entire distribution of currents flowing
in the Meissner-state part of the film was neglected. Moreover, the 
electric field is not considered in the simple analysis, and is also not 
available from the experiment. 

To complete the analysis of the flux guidance, we have developed an efficient numerical scheme allowing
to carry out simulations of the electrodynamic response of a thin
superconductor with antidots.  Below is a description of the scheme
followed by a report of the numerical results for a film patterned
just like the present YBCO sample.

\section{Simulation scheme}
 
The numerical scheme assumes that the superconductor is thin, i.e.,  it has a
 thickness $d$ much less than any lateral dimension of the sample. 
The external field, $H_a$, is applied in the perpendicular $z$-direction, and 
the induced currents will be quantified by the sheet current, $\mathbf J$. 
When $\mathbf J$ approaches
the critical magnitude $J_c=dj_c$, the depinning 
transition is sharp, and gives rise to a highly nonlinear
material characteristics conventionally approximated by a power
law,
\begin{equation}
  \mathbf E = \rho \mathbf J/d,\quad\rho=\rho_0 \left(J/J_c \right)^{n-1}
  ,
  \label{power-law-Ej}
\end{equation}
where $\mathbf E$ is the electric field and $\rho$ is the resistivity with $\rho_0$ being a characteristic value.
The creep exponent $n$ is usually large, typically in the range $20< n <60$ for YBCO.\cite{zeldov90,sun91} 
The Bean model corresponds to the limit $n\to \infty$.

Rather than working directly with the sheet current, 
a more convenient quantity is the local magnetization $g=g(\mathbf r,t)$,
defined by\cite{brandt95}
\begin{equation}
  \label{defg}
  \frac{\partial g}{\partial y}=J_x
  ,
  ~~~  
  \frac{\partial g}{\partial x}=-J_y
  ,
\end{equation}
where $\mathbf r = (x,y)$. The $g$ incorporates current conservation
since it by definition gives $\nabla \cdot \mathbf J = 0$.
The function $g$ is extended to the whole space by setting $g = 0$ outside 
the sample. Inside the antidots $g$ is uniform, but not necessarily zero. 

Another basic equation is the Biot-Savart law, which can be expressed as
\begin{equation}
  \label{b-s}
  B_z/\mu_0=H_a + \hat Qg
  ,
\end{equation}
with the operator $\hat Q$ given by\cite{vestgarden11}
\begin{equation}
  \label{hatQ}
  \hat Qg(\mathbf r) 
  = {\mathcal F}^{-1}\left[\frac{k}{2}\mathcal F\left[g(\mathbf r)\right]\right]
  ,
\end{equation}
where  $\mathcal F$ is the 2D spatial Fourier transform, and $k=|\mathbf k|$. 
The inverse relation is 
\begin{equation}
  \label{hatinvQ}
  \hat Q^{-1}\varphi(\mathbf r) 
  = {\mathcal F}^{-1}\left[\frac{2}{k}\mathcal F\left[\varphi(\mathbf r)\right]\right]
  ,
\end{equation}
where $\varphi$ is an auxiliary function. 

By taking the time derivative of Eq.~\eqref{b-s}, we get
\begin{equation}
  \label{dotg}
  \dot g = \hat Q^{-1}\left[\dot B_z/\mu_0- \dot H_a\right]
  ,
\end{equation}
which is solved by discrete integration forward in time.
This is possible since $\dot B_z(\mathbf r,t)$ can be calculated from $g(\mathbf r,t)$, 
as described below.

To solve Eq.~\eqref{dotg} the space is discretized 
in such way that $\hat Q$ and its inverse can be implemented 
using Fast Fourier Transforms. We let  the superconductor 
occupy the space $|x|<w$ and $y<|w|$, and 
by using periodic boundary conditions along $y$, the sample has the 
shape of a long strip in the $y$-direction. In the $x$-direction we let the strip be 
surrounded by empty space so that the total area included in the calculations
is  $|x|< 1.5w $, $y<|w|$.

Hence, the $xy$-plane consists of three different parts: 
 the superconductor,
 the area outside the strip, and
 the area inside the antidots. 
In order to solve Eq.~\eqref{dotg}, 
we must find $\dot B_z$ in all three regions,
which requires different algorithms. 

\begin{figure*}
  \centering 
  \includegraphics[width=16cm]{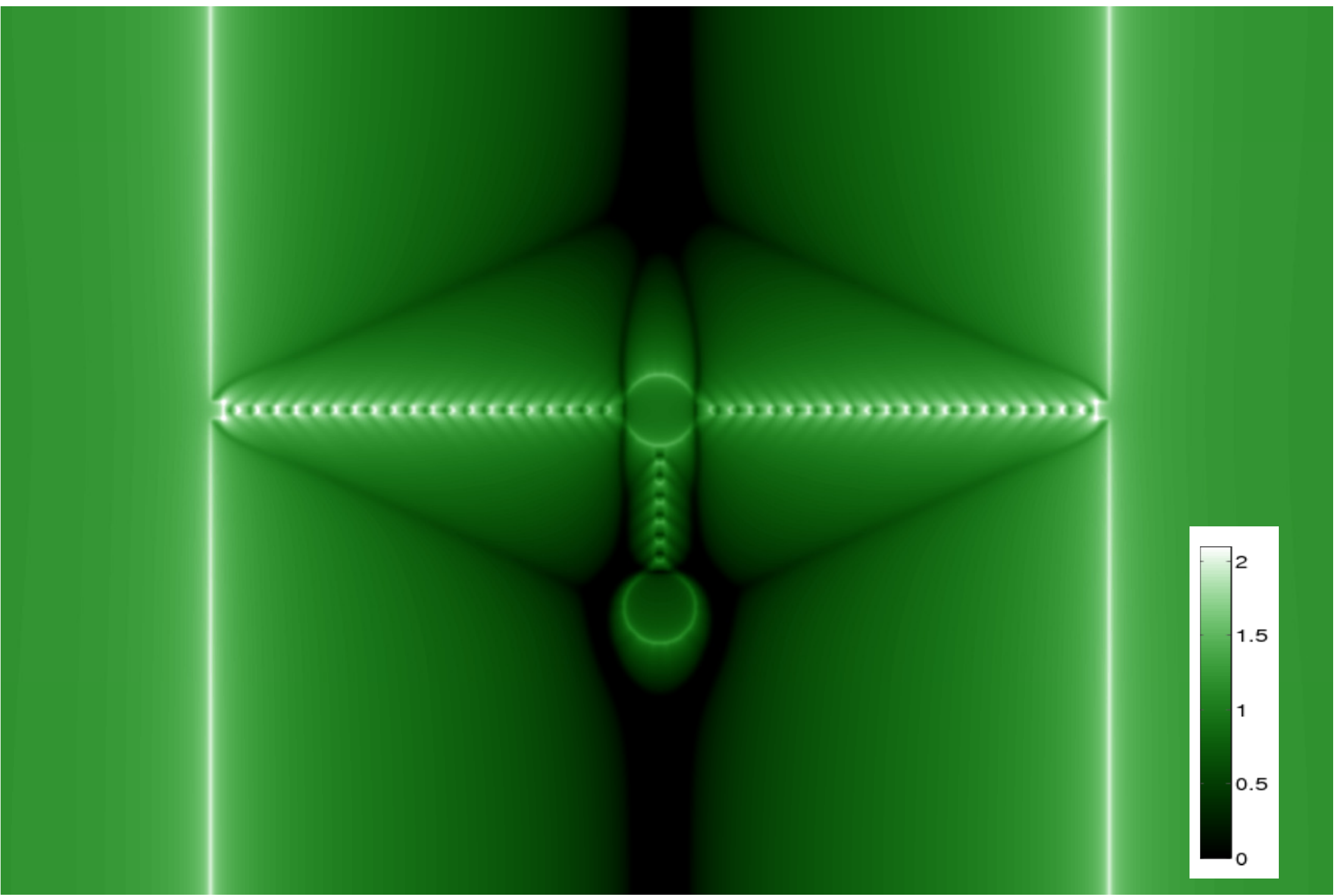}  
  \caption{
    \label{fig:full}
    (Color online)
    Numerical calculation of the distribution of $B_z$ in a
    superconducting thin strip after an applied magnetic field was
    increased from zero to $H_a=1$. The strip contains a pattern of
    antidots essentially identical to the sample studied
    experimentally. As in MOI the image
    brightness represents the magnitude of $B_z$ as indicated in the color bar insert.
  }
\end{figure*}

Starting with the superconductor itself,
it obeys the material law,  
Eq.~\eqref{power-law-Ej}, which when combined with
Faraday's law, $\dot B_z=- (\nabla \times \mathbf{E})_z$, 
gives
\begin{equation}
  \label{se2}
  \dot B_z = \nabla \cdot (\rho\nabla g)/d\, .
\end{equation}
From $g(\mathbf r,t)$ the gradient is readily calculated, and since
the result allows finding $\mathbf J(\mathbf r,t)$ from
Eq.~\eqref{defg}, also $\rho(\mathbf r,t)$ is determined from
Eq.~\eqref{power-law-Ej}. The task then is to find 
$\dot{B}_z$ in the non-superconducting parts, so that 
$\dot g=0$ outside the strip and  $\dot g$ is uniform
in the antidots. This cannot be calculated efficiently using direct 
methods due to the nonlocal relation between $\dot{B}_z$ and $\dot{g}$.  
Instead we use an iterative procedure.

For all iteration steps, $i=1...s$, $\dot{B}_z^{(i)}$ is fixed inside the
superconductor by Eq.~\eqref{se2}. At $i=1$, an initial guess is made for $\dot{B}_z^{(i)}$ 
outside the sample and inside the antidots, and $\dot g^{(i)}$ is
calculated from Eq.~\eqref{dotg}.  In general, this $\dot{g}^{(i)}$ does
not vanish outside the strip and an improvement is obtained by
\begin{equation}
  \label{iterative-Bz}
  \dot{B}_z^{(i+1)} = \dot{B}_z^{(i)}  -\mu_0\hat{Q}\hat{O} \dot{g}^{(i)} +C^{(i)} 
  .
\end{equation}
The projection operator $\hat O$ is unity outside the
strip and zero everywhere else. Also, the output of the operation should 
be shifted to satisfy $\int d^2r\hat Og=0$.
The constant $C^{(i)}$ is
determined by requiring flux conservation,
\begin{equation}
  \label{defC1}
  \int d^2r\, [\dot B_z^{(i+1)}-\mu_0 \dot H_a]=0  
  .
\end{equation}
Correspondingly, ${\dot B}_z^{(i+1)}$ in each of the antidots is also found by 
Eq.~\eqref{iterative-Bz}, but where the projection operator $\hat O$ now is unity 
in the antidot and zero everywhere else, and with
$C^{(i)}$ calculated using Faraday's law,
\begin{equation}
  \label{faraday}
  \int_\text{antidot boundary} \mathbf {dl} \cdot \mathbf E 
= -\int_\text{antidot area} d^2r \dot B_z
  .
\end{equation}

Thus, at each iteration $(i)$, we run through all antidots and
the outside area, and calculate $\dot B_z^{(i+1)}$. The procedure is repeated 
until after $i=s$ iterations
$\dot g^{(s)}$ becomes sufficiently uniform both outside the strip 
and within the antidots. Then, $\dot g^{(s)}$ is inserted in Eq.~\eqref{dotg},
which brings us to the next time step, where the whole iterative procedure 
starts anew.

Several comments can be made regarding the implementation 
of the simulation scheme.
First, the algorithm for finding $\dot g$ scales as $O(N\log N)$ with
the total number of discrete grid points $N$. This enables 
simulations of large grids, say $N > 10^6$.
Second, the quality and performance 
is improved by a good initial value, such as $\dot
B_z^{(1)}(t)=\dot B_z^{(s)}(t-\Delta t)$.  
Third, for small antidots there is a large perfomance gain by
replacing the full operator $\hat Q\hat O$ in Eq.~\eqref{iterative-Bz} by a
local operator that only runs over the antidot.
Fourth, the results can be made more robust by enforcing $\dot J=0$
exactly by hand after each iteration step to prevent accumulations of
small unphysical currents inside the antidots with time. 
Finally, a robust way to satisfy Eq.~\eqref{faraday} is to first
calculate $\dot B_z$ from Eq.~\eqref{se2} in the antidot area using the
material law as if the antidot was not there, and then choose $C^{(i)}$
so that the total flux in the antidot is the same. 

At all times, the simulation scheme provides direct access to the distributions of 
$g$, $B_z$, $\mathbf J$, everywhere in the plane $z=0$, as well as access to $\mathbf E$ 
inside the superconductor, through Eq.~\eqref{power-law-Ej}. Note that $\mathbf E$ is 
not calculated inside the antidots or outside the sample.

\section{Simulation Results}

Numerical simulations were carried out for a superconducting 
strip  of width $2w$ where between the two edges there is a linear
array of 40 antidots of radius $r= 0.01w$ and center-to-center
separation $a = 4.4r$.  Two larger holes of radius $0.08w$ are
located on the strip center line, and with another array of 6 antidots
in between. The area of calculation, $3w \times 2w$,  was discretized on a 
$1536\times 1024$ equidistant grid. The creep exponent was $n=29$.
The simulation was carried out in dimensionless units based on 
$J_c'=J_c (wd\mu_0\dot H_a/\rho_0J_c)^{1/n}$. The results can be converted 
to dimensional units by the transformations $J\to JJ_c'$, $B_z\to B_z\mu_0J_c'$,
$t\to tJ_c'/\dot H_a $ and $E\to E\rho_0J_c'/d$.

Shown in Fig.~\ref{fig:full} is the result of calculating the
distribution of $B_z$ at an applied field of $H_a=1$ during a field
ramp at the rate $\dot H_a=1$ starting from $H_a=0$.  The magnetic
flux penetrates from the edges where a critical state is established
with $J\approx 1$. The antidots are clearly seen, and the
flux distribution reproduces all characteristics of the
magneto-optical image of Fig.~\ref{fig:mo}.  In particular, the
simulation reproduces the high $B_z$ along the array of antidots,
i.e., the flux guidance provided by the patterning.  The excellent
qualitative agreement with the experimental image on all visible
scales gives strong confidence in the correctness of the simulation
method. 

Also the d-line$_1$ and d-line$_2$, which are clearly seen in Fig.~\ref{fig:full},
are in accordance with both the experiment and the Bean model
considerations.  Interestingly,  the dark lines d-line$_1$ make an angle 
$\alpha = 66^\circ$ with the strip edges, which is slightly larger than 
expected from the Bean model, where Eq.~\eqref{alpha} gives $\alpha = 62^\circ$
for $a/r=4.4$. This implies
that the current flow is enhanced through the bridges between the antidots 
due to the use of a finite creep exponent, $n$.

\begin{figure}
  \centering
  \includegraphics[width=\columnwidth]{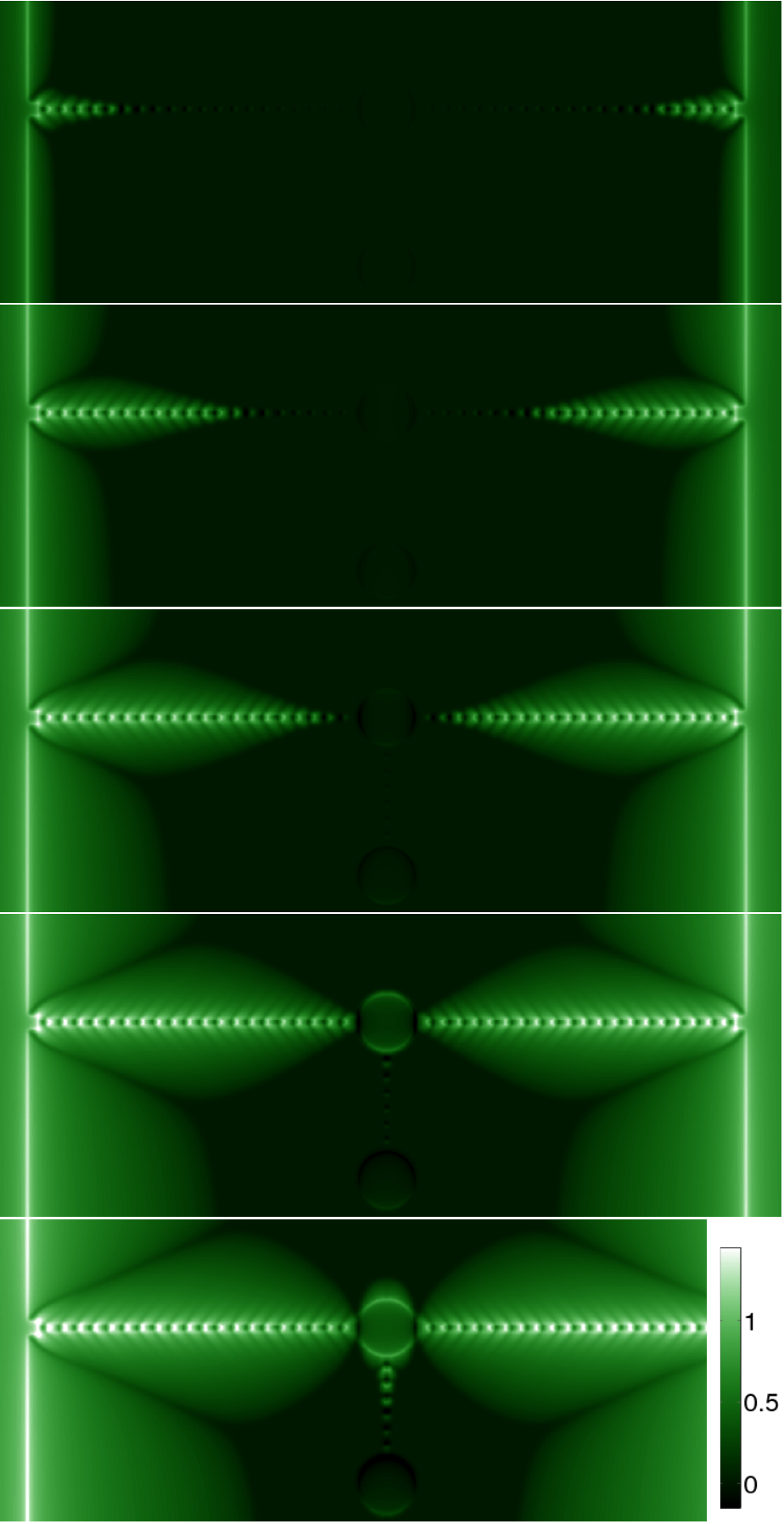}  
  \caption{
    \label{fig:sim1234}
    (Color online)
    Distribution of $B_z$ in the patterned superconducting strip at  
    increasing applied fields of $H_a=0.1$(upper),  $0.2$, $0.3$, 0.4 and $0.5$(lower). 
The inserted color bar relates the local magnitude of $B_z$ and the image brightness. 
  }
\end{figure}

Figure~\ref{fig:sim1234} illustrates how the flux distribution evolves
as the applied field is ramped up. The five panels show simulated
images from $H_a=0.1$ to 0.5.  Evidently, the flux penetration is at
all stages greatly advanced along the antidot array. In fact, it extends even 
deeper than expected from the Bean model considerations in Fig.~\eqref{fig:bean}.
This is due to the currents flowing in the Meissner
state part of the film, where they reach the critical value when adapting to the 
constrictions created by the antidots. 

At the field $H_a=0.4$, a continuous area with non-zero $B_z$ connects
the edges and the central large hole, which first at this stage
receives sizable amounts of flux.  Beyond that stage, we find both
experimentally and numerically that one may tune the amount of flux
captured in the hole by increasing or decreasing $H_a$, and making the
hole act as a controllable flux reservoir.  Note that at $H_a=0.4$ the
second large hole is not yet in contact with the flux front, but still
faintly visible in the figure due to its perturbation of the Meissner
current flow.  In the final panel, $H_a=0.5$, flux is guided from the
first large hole towards the one below following the connecting array
of antidots.  Here, the flux motion is parallel to the edges. The
whole sequence of flux penetration patterns during increasing applied
field agrees very well with the experimental results shown in
Fig.~\ref{fig:mo}.

\begin{figure}[t]
  \centering
  \includegraphics[width=\columnwidth]{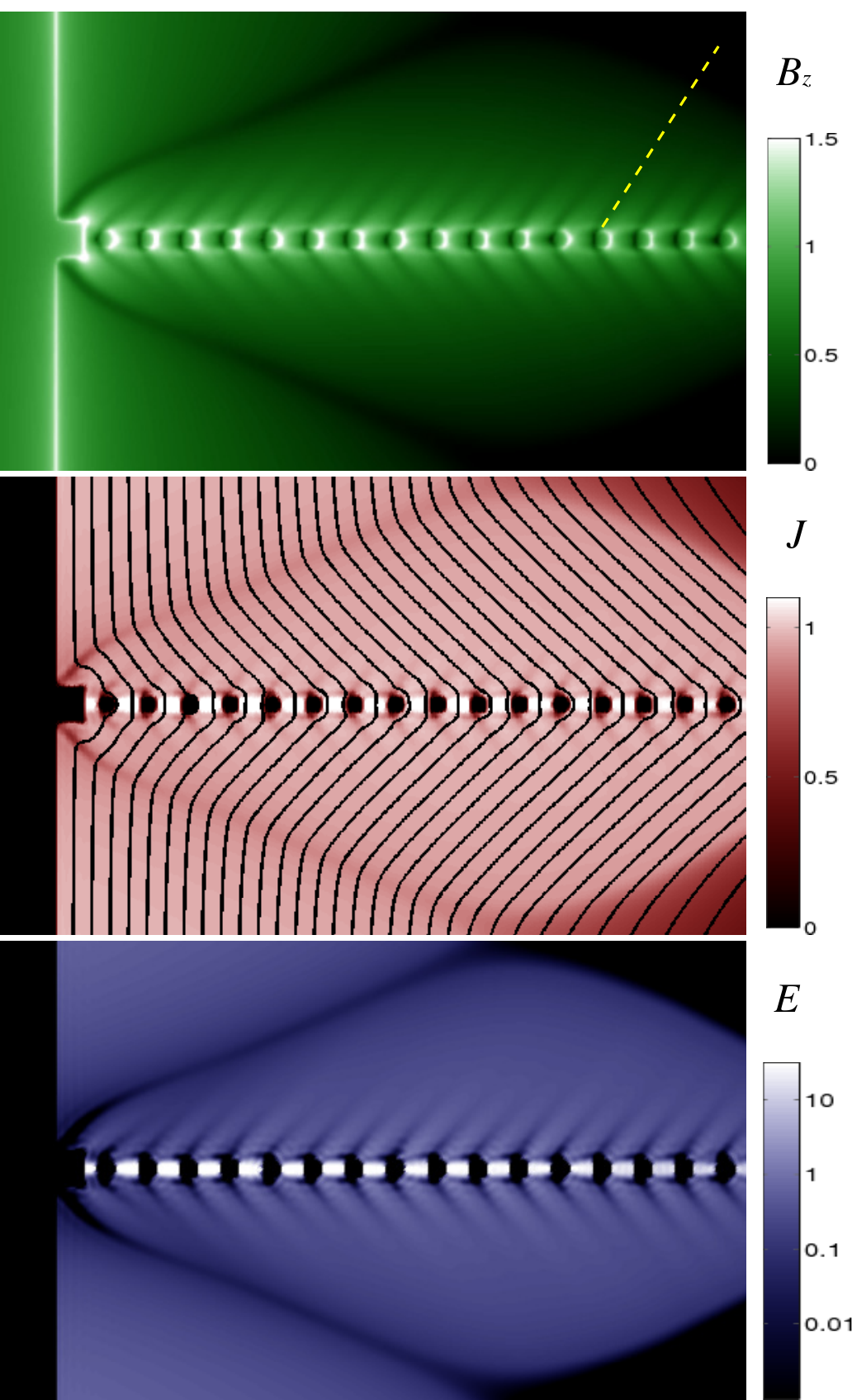} 
  \caption{
    (Color online)
    Distributions of $B_z$, $J$ and $E$ along 
    the left antidot array, at $H_a=0.4$.
    The $J$-map is superimposed with the current stream lines.
    In the $E$-map, the white spots are the bridges 
    between the antidots, where the flux traffic is 
    extensive. Color bars relating the image brightness and local values are shown. 
    \label{fig:BandE}    
  }
\end{figure}

\begin{figure}[t]
  \centering
  \includegraphics[width=\columnwidth]{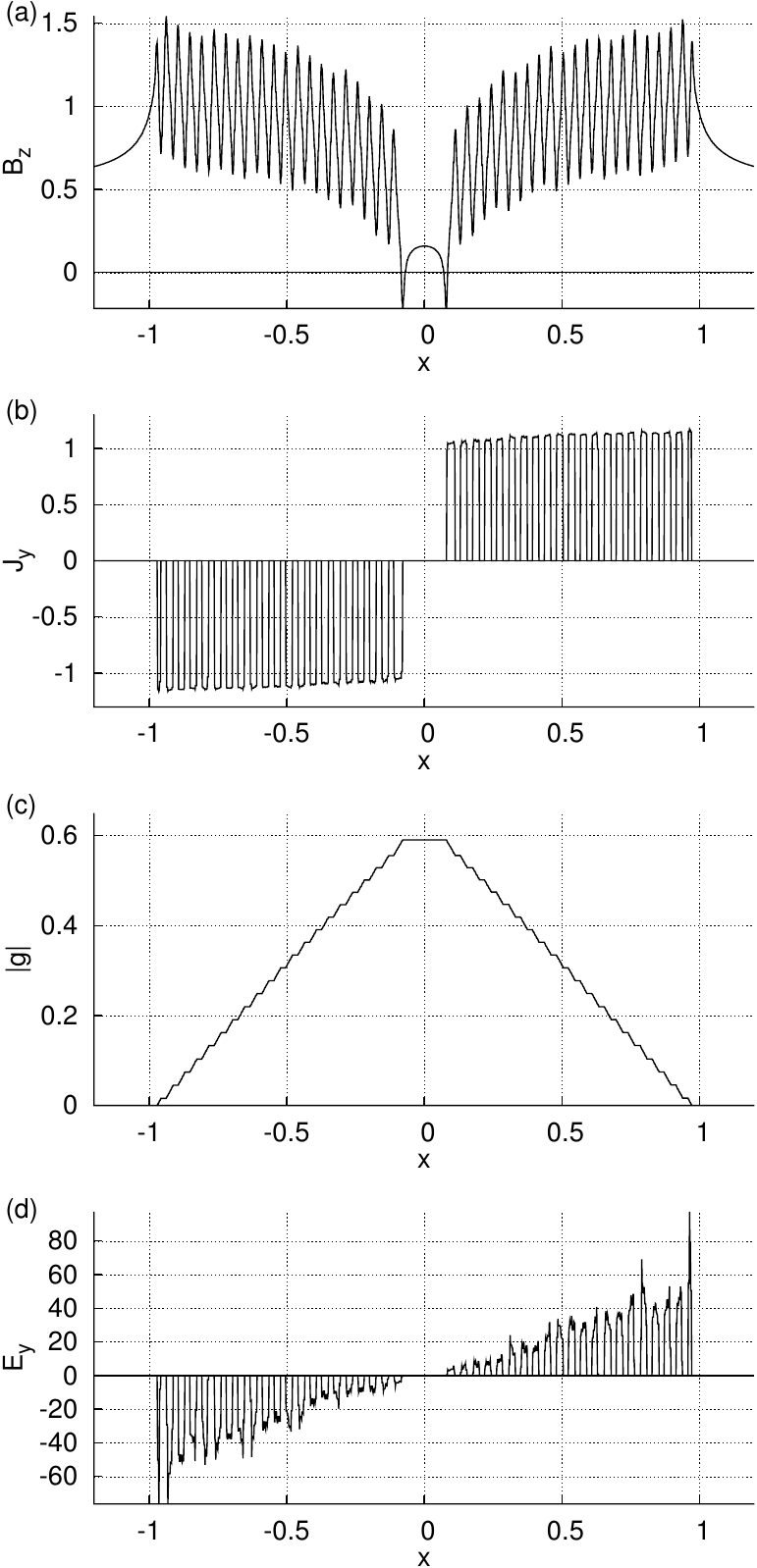} 
  \caption{
    \label{fig:profiles}
    Profiles of (a) $B_z$, (b) $J_y$, (c) $|g|$, and (d) $E_y$ across the antidots at $H_a=0.4$.}
\end{figure}

Figure~\ref{fig:BandE} presents a close-up view of the antidot array 
at $H_a=0.4$. The upper panel shows the $B_z$, where now the fishbone 
structure is seen. The Bean model result,  Eq.~\eqref{beta},
predicts for the present array that the d-line$_2$ forms the angle
$\beta = 33^{\circ}$ with the edge. A  dotted line at this
$\beta$ is included as a guide to the eye in the figure, and it demonstrates
very good agreement. 

The middle panel shows the magnitude of the sheet current $J$ with the 
current stream lines superimposed. The main features of the current stream lines are in excellent 
agreement with the Bean model construction of 
Fig.~\ref{fig:bean}, but the curving of the stream lines is less sharp due 
to the finite $n$. In particular, the cusps creating the d-line$_2$ are  
weak. Note that some current stream lines extend into the flux-free 
Meissner state region, as expected in films placed in a perpendicular magnetic field.

The lower panel shows the magnitude of the electrical field
$E=|\mathbf{E}|$ in the same area. To reveal the overall field
distribution the images has been plotted on a logarithmic scale, with
$E$-values ranging from $10^{-3}$ to 30.  Very large fields, $E > 30$,
were found in the bridges of superconductor connecting the
antidots. Evidently, the antidot array forms a channel with large
traffic of magnetic flux.  At the same time, the flux motion is much
reduced at the d-lines, which show up as dark also in the map of $E$,
indicating suppressed traffic of magnetic flux. This is a generic
feature of regions with sharply curved current
streamlines.\cite{brandt95}

A quantitative presentation of the profiles for $B_z$, $J$, $g$, and
$E$ across the antidots array at $H_a=0.4$ is shown in
Fig.~\ref{fig:profiles}. Clearly, all the quantities are much
distorted compared to those of an unpatterned strip.\cite{norris69,
brandt93, zeldov94} In particular, $B_z$ has an oscillating behaviour
with minima at the inner, and maxima at the outer, edges of the
antidots, where the peak values are comparable to those along the
strip edge. The sheet current $J_y$ is zero within the antidots and is
almost constant in between, with values $J\approx 1.1$.  The the local
magnetization $g$ has the shape of a step pyramid with flat levels
across the antidots.  The maximum electric field is very high, $E>
50$, compared to a plain strip where $E<1$ at the edge, and it is also
higher than for small indentations at a film edge.\cite{schuster96,
vestgarden07} The high $E$-value reflects that all magnetic flux in
the diamond-shaped region has passed through channels of width
comparable to the antidot diameter.  Numerically, the values $E\approx
50$ and $J\approx 1.1$ are consistent, since from
Eq.~\eqref{power-law-Ej}, one has $50^{1/n}\approx 1.14$ for $n=29$.

Previous works concerned with the stability of superconducting films have found that 
high electric field close to the edge implies that the sample is
susceptible for avalanches triggered by thermomagnetic instabilities.\cite{mints96,denisov05}
This means that a transverse array of antidots, or small defects, 
are likely nucleation points for the instabilities. 
In thin films, the consequences of the instabilities are dramatic, as they
often take the form of large dendritic structures as observed in many
materials.\cite{Duran95,bolz03, rudnev05, choi05}

Finally, note that the results presented in Section II were based only 
on current conservation, and will give a quite good description for 
any sample thickness provided the superconductor behaves according 
to the critical-state model.
Thus, also the general concept of flux guidance as presented in this work 
should be essentially independent of thickness. 
On the other hand, the simulation formalism, which
gives a more precise description, was derived under the assumption 
that the thickness is much smaller than the size of any lateral structure
in the sample. Therefore, we expect that the presented simulation results 
will hold only as long as the antidot diameter is considerably larger 
than the film thickness, as was the case in the present experiments.

\section{Summary}

Magnetic flux guidance by linear arrays of antidots in type-II superconducting
films has been considered experimentally and theoretically.  
Experimentally we have used MOI to show strong flux guidance in a YBCO film
shaped as a long strip and patterned with antidots arranged in linear arrays.
It was also shown that an antidot array along the center line can promote flux
motion parallel to the strip edge.
The flux penetration patterns have revealed that the antidot arrays
perturb the overall flow of current considerably. 
In particular, we find that lines where the current flow abruptly changes
direction, the d-lines, agree very well with current stream line
patterns constructed from the Bean critical-state model.  Further
insight into the flux penetration process was achieved by numerical
simulations of the electrodynamic response of the superconductor
subjected to an increasing perpendicular magnetic field.

In order to perform simulations of superconducting films with
complicated antidot patterns it was necessary to develop a efficient
method for imposing boundary conditions, both for the sample
boundaries and the antidots.  The simulation of an ascending field
ramp produced a flux distribution with the same qualitative and
quantitative characteristics as found in the experiment. In addition,
the simulations give a deeper insight into the dynamical process of
flux guidance by providing information of all electrodynamic
quantities at all stages of the process. It was shown that the
electric field becomes very large in a thin channel connecting
the antidots, in particular close to the edges. 
Also, maps of the electric field show that flux motion is
much suppressed elsewhere near the antidots. This means that
magnetic flux is guided into the film via a main route along the
antidot array.

\acknowledgments
The work was supported financially by the Norwegian
Research Council.

\bibliography{superconductor}

\end{document}